\documentclass[12pt]{article} 

\usepackage{epsf} 
\usepackage{graphicx}
\usepackage{amssymb}
\usepackage{amsmath}
\usepackage{epstopdf}
\usepackage{xcolor}

\newcommand{\beq}{\begin{equation}} 
\newcommand{\eeq}{\end{equation}} 
\newcommand{\beqa}{\begin{eqnarray}} 
\newcommand{\eeqa}{\end{eqnarray}}

\begin{document}

\bigskip\bigskip\bigskip 
 
\begin{center} 

{\Large\bf On the Topic of Emergence from an\\[0.3cm] Effective Field Theory Perspective}

\end{center} 
 
\vspace{.2in} 
 
\begin{center} 
{\large  Thomas Luu$^{1,2}$~\footnote{E-mail: t.luu@fz-juelich.de}~and~
Ulf-G. Mei{\ss}ner$^{2,1,3,4}$~\footnote{E-mail: meissner@hiskp.uni-bonn.de}}

\bigskip 
 
\bigskip 
 
$^1$~{\it Institute for Advanced Simulation (IAS-4), Institut f\"ur Kernphysik (IKP-3),\\
and J\"ulich Center for Hadron Physics, Forschungszentrum J\"ulich, Germany} 
 
$^2$~{\it Helmholtz-Institut f\"ur Strahlen- und Kernphysik and Bethe Center for Theoretical Physics, Rheinische Friedrich-Williams-Universit\"at Bonn, Germany}

$^3$  {\it Center for Science and Thought, Rheinische Friedrich-Williams-Universit\"at Bonn, Germany}

$^4$~{\it Tbilisi State University, 0186 Tbilisi, Georgia}

\end{center} 
 
\vspace{.7in} 
 
\thispagestyle{empty}  
 
\begin{abstract} 
\noindent
Effective Field Theories have been used successfully to provide a ``bottom-up" description of phenomena whose
intrinsic degrees of freedom behave at length scales far different from their effective degrees of freedom.
An example is the emergent phenomenon of bound nuclei, whose constituents are neutrons and protons, which
in turn are themselves composed of more fundamental particles called quarks and gluons.  In going from a
fundamental description that utilizes quarks and gluons to an effective field theory description of nuclei,
the length scales traversed span at least two orders of magnitude.   In this article we provide an
Effective Field Theory viewpoint on the topic of emergence, arguing on the side of reductionism and weak
emergence.   We comment on Anderson's interpretation of constructionism and its connection to strong emergence. 
\end{abstract} 
 
\vspace{1.3in} 
 
 
 
\vfill

\newpage 
 
\section{Emergent phenomena in  nuclear physics}

The term \emph{emergence} is a multi-faceted concept whose exact meaning depends on context and invariably
the field of study.  In the field of (low-energy) nuclear physics, emergent phenomena are always associated
with highly complex and highly non-linear behavior.  Such phenomena are deemed \emph{non-perturbative}; their
descriptions at the fundamental (lower) level are not amenable to simple paper and pencil calculations.
An example, which we will go into more detail later, is the behavior of quarks and gluons (the fundamental
particles at the lower level) and how they come together to form protons and neutrons, or collectively nucleons.
The theory of the interactions of these lower level particles is known as quantum chromodynamics (QCD). It is
a seemingly simple theory which can be written down in one line. However, the manner in which three quarks
interact (via the exchange of gluons) and thus bind themselves to form nucleons is the quintessential
non-perturbative problem. The explanation of the observation that  quarks and gluons are never seen as free particles
but are rather confined within strongly interacting particles, the hadrons such as the proton and the neutron,
constitutes one of the grand challenges in theoretical physics~\cite{clay}.
Complicating matters is the fact that gluons, which are considered the force carrier~\footnote{The photon is
the analogous force carrier for quantum electrodynamics (QED).} of the ``strong interaction" between quarks,
can also interact with themselves.  This has profound implications on the generation of mass.  Gluons are massless,
and while quarks do have mass, their masses are roughly two orders of magnitude smaller than the mass of a
nucleon.  Thus most of the mass of the nucleon does not come from the masses of its (matter) constituents; rather, it
is generated dynamically by the interactions of the gluons.  This extends to the elements built from nucleons.
Indeed, ninety-five percent of the mass of the \emph{observable} universe is generated from interactions
between \emph{massless} particles (gluons).  The term holistic might be an understatement in this case.

Emergent nucleons from bound quarks subsequently bind and form heavier elements, such as deuterium, helium,
carbon, oxygen, and so on.  These processes are themselves emergent and non-perturbative.  And the phenomena
that emerge from such elements are vast and complex.  Nuclear breathing modes for example, where large groups
of nucleons within a nucleus move in a collective motion, lead to oscillatory behavior called giant dipole
resonances.  Such resonances, when coupled with electromagnetism and the weak interaction, play an integral
role in nuclear fission.

Another example and one that is not unique to nuclear physics, involves closed three-body systems known as
Borromean states. The constituents of these states, when only considered pairwise, are not bound.  However,
when a third constituent is included, the system becomes bound and the resulting spectrum is extremely
rich and diverse.  Examples of nuclear Borromean states are $^6$He, whose constituents are $^4$He and two
neutrons and the Hoyle state of carbon, whose constituents in the case are three separate $^4$He nuclei~\footnote{
Such states are only truly Borromean if we neglect the weak and electromagnetic interactions.}.

The examples of emergent phenomena above have little to no resemblance to their lower level constituents,
which in this case are quarks and gluons.  And since their description at the lower level via traditional
calculations is essentially all but impossible, physicists instead turn to the powerful tool of
\emph{effective field theory}, where instead of using the lower level constituents to frame the problem,
they instead work directly with the emergent phenomena as the relevant degrees of freedom~\footnote{We remark that
  in an EFT, the lower-level theory indeed acts at higher energies and vice versa.}.  This effective
field theory (EFT) is not an \emph{ad hoc} description of the emergent phenomena, however.  If
developed properly, the EFT represents an equally valid representation of the phenomenon and can be used
to predict new phenomena as well as to \emph{verify the lower level theory}. We remark in passing that
nowadays it is widely accepted that all field theories are effective field theories, which makes the
phenomenon of emergence even more ``natural''.

In the following section we give a cursory primer on effective field theories whereby we enumerate the
ingredients and conditions for constructing a successful EFT.  We give examples of EFTs both from a
historical point of view and from a modern viewpoint.  In Sec.~\ref{sect:EFT and emergence} we discuss the
relationship between EFT and Emergence/Causation.  We pay special attention to both strong and weak forms and argue on the side of weak emergence and reductionism.  In Sec.~\ref{sect:more is different} we provide our own interpretation of Anderson's thesis in his
seminal paper ``More is different'' \cite{Anderson:1972pca}, whereby we compare with Ellis' interpretation given in \cite{ellis2016}.  We continue our discussion in Sec.~\ref{sect:purpose} on the apparent dichotomy of physics and biology due to ``purpose", as motivated by Ellis in \cite{ellis2016}, but argue that such reasoning is false and misleading.   In Sec.~\ref{sect:popper}  we invoke Popper's falsifiability argument to gauge
the ``scientific merit" of strong emergence.  Finally, we recapitulate our arguments
in Sec.~\ref{sect:conc}.

\section{Primer on \emph{effective field theory}}
Since emergent phenomena preclude simple calculations from their constituent basis, physicists instead use
\emph{effective} degrees of freedom to describe these systems. In EFT terms, these are dubbed the ``relevant
degrees of freedom'', as will become clearer later on. Nucleons (protons and neutrons), rather than quarks and
gluons, for example, form the basis for describing nuclear phenomena.  The interactions between nucleons are
not disconnected from the interactions of quarks and gluons, however.  They are related to their constituent
parts in a rigorous, systematic manner.  This procedure of relating the effective degrees of freedom to the
dynamics of constituent parts collectively falls under the purview of effective field theory. We discuss
tersely the necessary and sufficient ingredients for constructing an effective field theory below. These are:

{\bf Identification of \emph{effective}, or \emph{active degrees of freedom}:}~ Here the emergent phenomena
(e.g. protons, pions, nuclei, breathing modes in large nuclei, etc. . .) dictate the active (relevant) degrees
of freedom, despite the fact that such phenomena can be expressed as collections of more fundamental degrees of
freedom (i.e. constituents).  The energy scales at which the emergent phenomena operate are considered
\emph{low} compared to the intrinsic energy scales of its constituents. This is very obvious in nuclear
physics, while nuclear excitations involve energies of say tenth of MeV, to investigate (see) the quark-gluon substructure
of nucleons or nuclei requires probes with multi-GeV energies, i.e. three orders of magnitudes larger.
This identification of the relevant degrees of freedom can also be understood from the principle of
resolution: The finer details of a system one wants to investigate, the larger energy (momentum) is needed, as
follows simply from Heisenberg's uncertainty principle. Therefore, as stated before, energies of
relevance to nuclei can never reveal their quark-gluon substructure, it is simply irrelevant at these
energies.

{\bf Separation of length scales:}~ The separation of length (or: energy) scales is implicit in all EFTs.
By their very definition, emergent phenomena occur at length scales that are larger than their constituents'
intrinsic scales.   Put another way, the energy required to resolve the emergent phenomenon is insufficient
to resolve its constituents.  Such separation in length scales, or equivalently energy scales, allows
one to express an EFT as an expansion in the ratio of scales (e.g. the ratio of the energy scale of
the emergent phenomenon to its constituents' intrinsic energy scales).  The larger the separation of scales,
the more \emph{effective} the EFT description as such an expansion converges faster. 

{\bf Identification of symmetries:}~Symmetries play a fundamental role in the construction of any EFT of
some emergent process.  The symmetries that the emergent phenomenon respects are identical to the
symmetries of its constituents and their interactions.  Examples of such symmetries are Lorentz invariance
(physics does not depend on the observer's frame), gauge invariance (local transformations of quantum fields
that leave the Lagrangian invariant), and the discrete
symmetries of parity (e.g. mirror symmetry), time-reversal invariance (a reaction happening forward in time
is equal to its counterpart happening  backward in time), and charge
conjugation (particles turn into anti-particles).  The identification of such symmetries provides strong
constraints in the types of expressions that show up in an EFT, and in many cases, simplifies and
reduces the number of terms by providing relations between different expressions.
Despite these constraints on the form of terms, each term has an associated coefficient that is not
determined by symmetry alone and must be either empirically determined or derived from the lower level theory.
It should also be noted that symmetries can be {\em realized} differently in the EFT compared to its underlying
theory or can be broken upon quantization. Such phenomena happen indeed in QCD and its corresponding EFT,
but we will not discuss them any further here.

{\bf Power counting scheme:}~Even with the identification of all terms with the relevant symmetries of the
system in question, there still exists myriads of terms~\footnote{In principle there are an infinite number of
  expressions.} that make any EFT calculation futile unless there is some systematic way of organizing the
expressions in terms of relative importance. Here one employs the concept of power counting, where the different
terms are enumerated in hierarchical importance related to some expansion parameter (usually related to the
ratio of some soft momentum scale to a hard scale).  A desired accuracy of calculation then dictates the
number of terms to be calculated. Thus there is a finite number of terms that are needed for any EFT calculation.
Note, however, the following ramification of such a procedure:  no EFT calculation will ever be exact since there
is always an associated uncertainty (due to the truncation of terms).  Also, it should be noted that any realistic
calculation in physics can never be exact (unless in very simplified toy models).

\subsection{Examples of Effective Field Theories}
Here we give a few examples of well known EFTs.  These examples are by no means exhaustive, they are simply
selected because of their inherent clarity.\\

{\bf Heisenberg-Euler Theory:}~ In the early 1930s Heisenberg, Euler and Kockel considered
a theory~\cite{Heisenberg:1935qt,Euler:1935zz}
describing photons with energies $\omega$ that are much lower than the mass of the electron, i.e.
$\omega\ll m_e\ (=.511\  \mathrm{MeV})$~\footnote{1 MeV = 1 mega-electronvolt = $10^{6}\,$eV.}.
As there are no other electromagnetically charged particles with
masses less than the electron, the only active degrees of freedom are thus the low-energy photons.
All other heavier particles, here electrons and their anti-particles, have been \emph{integrated out}.
Their theory, constrained by Lorentz symmetry and parity, constitutes essentially an EFT expansion in
$(\omega/m_e)^2$ (odd powers are forbidden by parity), and was used to investigate the cross section for
light-by-light scattering at low energies.  Note that their effective theory allows for a direct interaction
between photons, whereas in the more fundamental theory, quantum electrodynamics (QED), a photon cannot
interact directly with another photon, but only through an intermediary charged particle. The corresponding
full calculation in QED including electrons as active degrees of freedom was only done in 1951~\cite{Karplus:1950zz},
and found to agree with the EFT result in the energy range $\omega \ll m_e$.

\begin{figure}
\center
\includegraphics[width=.6\textwidth]{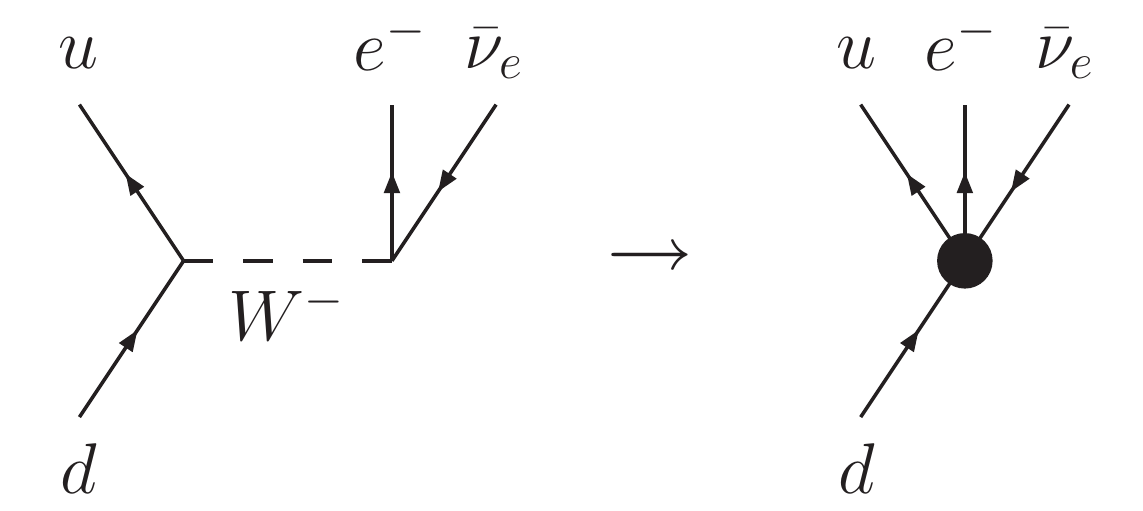}
\caption{Neutron decay as seen from the fundamental weak interaction theory, where a down (d) quark turns
into an up (u) quark with the emission of a heavy $W^-$-boson that then decays into an electron and an anti-neutrino.
At energies much below the $W$-mass, the propagation of this virtual particle can no longer be resolved and the
interaction collapses into a four-fermion point coupling. Not shown a the up and down quarks not participating
in the decay, as a neutron is made from quarks as $|udd\rangle$ and a proton as $|uud\rangle$.
\label{fig:weak}}
\end{figure}

{\bf Fermi's Theory of Beta Decay:}~ Also during this time Fermi proposed a theory~\cite{Fermi:1934hr} for
beta (weak) decays that utilized a
direct four-fermion coupling (dubbed four-Fermi interaction later).  An example of such a process is the decay
of a neutron into a proton, an electron, and an anti-neutrino, $n\to p e^- \bar{\nu}_e$  The theory worked
amazingly well at low energies, but was shown to break down at energies of order $\sim100$~GeV.  In light
of our understanding today of the more fundamental theory of weak interactions, such a breakdown is not
surprising.  The particles that mediate the weak interaction are the $W$ and $Z$-bosons, whose masses
are approximately $M_W\sim80$~GeV.  At energies well below this mass, these particles cannot be resolved
(as discussed above) and Fermi's effective four-fermion coupling is an excellent approximation
to the weak processes, as shown in Fig.~\ref{fig:weak}.  But at energies comparable or larger to $M_W$,
these particles are now resolvable and the effective field theory breaks down.  At these high energies one
is forced to work with the more fundamental theory and its constituents.

{\bf Chiral Perturbation Theory:}~Despite knowing the more fundamental (lower level) theory of its constituents,
the above examples provide two effective theories that are just as good at describing, and \emph{predicting},
processes related to their emergent phenomena (i.e. light-by-light scattering and beta decay), as long as one
is willing to work at low enough energies.  Indeed, calculations with the effective theory are more often
than not simpler to perform in these low energies.  But given the theoretical advances in our understanding
of QED and the weak interaction, today's physicists prefer to work directly with the more fundamental
theory when it comes to photons and weak decay.  We stress, however, that either representations (EFT or
fundamental theory) are valid descriptions of these emergent phenomena.

\begin{figure}
\center
\includegraphics[width=.6\textwidth]{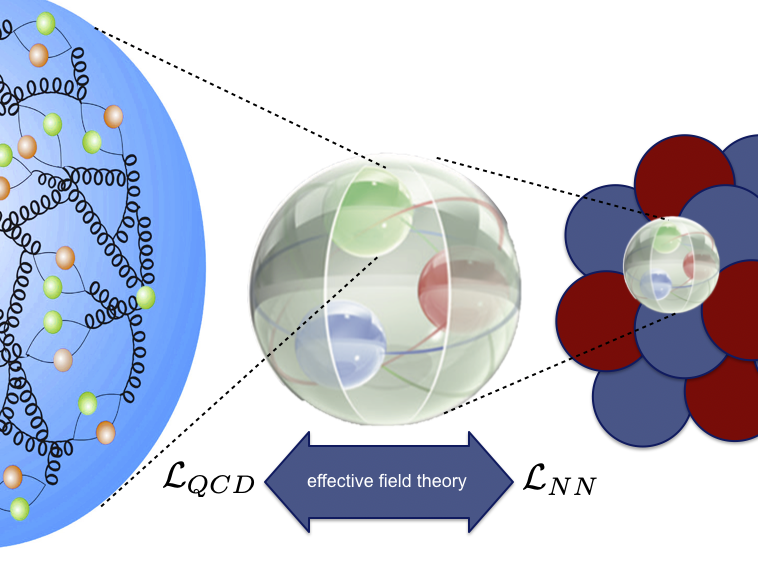}
\caption{Chiral perturbation theory provides an effective multi-nucleon interaction $\mathcal{L}_{NN}$ (right) that
is constrained by the symmetries and interactions $\mathcal{L}_{QCD}$ between the lower level, fundamental
quarks and gluons (left). Note the length scale traversed by these systems, ranging from sub-femtometers (left) to 10s of femtometers (right).  The size of the nucleon (center) is approximatelly 1 femtometer.  \label{fig:qcd_to_nn}}
\end{figure}

On the other hand, Chiral Perturbation Theory~\cite{Weinberg:1978kz,Gasser:1983yg} is an
EFT that serves as a more modern example where, despite knowing its underlying theory of Quantum
Chromodynamcs (QCD), one cannot directly perform calculations with the fundamental theory (at energies
$\lesssim 250$~MeV)~\footnote{Such calculations can be done on a finite volune space-time, known as lattice QCD,
 but this requires state-of-the-art supercomputers and will not be discussed further.}
but must utilize its corresponding EFT. More precisely, there are two distinct EFTs for QCD, one refers to
the sector the the light quarks (up, down and strange) and the other to the heavy quarks (charm and bottom).
In what follows, we will consider the the light quark sector only. In this sector, the active degrees of
freedom are nucleons (neutrons and protons) and pions, and the interaction between these degrees of freedom
originate from the interactions of their constituents, the light quarks and gluons.  It is important to note the length
scales encompassed by these systems, which range for sub-femto\footnote{One femto(meter) is $10^{-15}$ meter.} (quarks and gluons) to 10s of fermi (nuclei),
as depicted in \ref{fig:qcd_to_nn}.  The form of the interaction terms is dictated by a chiral symmetry
\footnote{Chiral symmerty refers to the fact that in massless QCD, one can write down two indepedent theories in
terms of left- and right-handed quarks, respectively. This symmetry is explicitely broken due to the
small quark masses.} of the quarks (in the massless limit), the coefficients of which are constrained by
empirical data.  The diverse separation of scales, coupled with a consistent power counting
scheme, provides the organizational tools to perform calculations that both postdict and
predict nuclear emergent phenomena~\cite{Epelbaum:2008ga}. Coupled with high-performance computing,
calculations of nuclei up to the mid-mass region ($\sim50$ nucleons) are now possible~\cite{Lahde:2019npb}.

{\bf The Standard Model of Particle Physics:}~ The Standard Model of Particle Physics (SM), which
encompasses the strong (QCD), weak, and electromagnetic (QED) interactions, is in and of itself
an incomplete theory since it does not include gravity.  We know that new physics must occur at least at
the Planck energy scale~\footnote{The Planck scales signifies the point where gravity and the SM forces
have equal strengths.}, $1.2\times 10^{25}$~MeV, and most likely before this.  Relative to this energy scale,
the standard model is itself a low-energy EFT.  This means that the QCD, weak, and QED interactions are
\emph{not exact}, but low-energy approximations of some grander theory.  Still, the accuracy of these
``effective" theories is very high due to the large separation to the Planck scale (or, more generally,
the scale of physics beyond the SM, which is estimated to be in the TeV region). 

\section{EFT in relation to \emph{Emergence} and \emph{Causation}
\label{sect:EFT and emergence}}

The EFT description provided above naturally leads to a \emph{bottom-up} approach, where upper-level
emergent phenomena and their associated larger length scales/lower energies are built from lower-level (more) fundamental constituents.   The level below the EFT is required to calculate 
   certain properties from more basic constituents, like e.g the values of the low-energy constants (LECs). This is fully consistent with
the \emph{reductionist} point of view.  The entire field of particle physics, whether it is consciously
aware of this or not, follows the reductionist line of reasoning, at least in \emph{methodological} and \emph{theoretical} forms:  we build experiments and theories
that probe lower level physics in an attempt to investigate currently unexplainable phenomena.  Such phenomena, if deemed ``emergent'',  serve as parametrizations of our ignorance.

It is natural to think then that causation follows this same bottom-up (or \emph{upward}) direction as well.  Indeed, within the
EFT prescription, it is the symmetries of the lower level that dictates the allowed interaction terms at
the higher level, but not the other way around.  However, as the EFT description is a fully consistent
and equally valid representation of emergent phenomena, any prediction it makes, regardless of how ``disconnected''
or ``unexpected'' when viewed from the lower level theory, is consistent with the laws that govern the
lower level constituents.  Such predictions, and the associated causal impacts that accompany them,
are in principle~\footnote{In practice, such a deduction might be impossible due to computational constraints.}
\emph{deducible} from the lower level constituents.  This is consistent with the notion of \emph{weak downward}
causation~\cite{doi:10.1111/0029-4624.31.s11.17}.  

When it comes to emergence, the same applies for EFTs.   In principle, calculations at the lower level using QCD would be preferable, all things being equal.  Indeed, the LECs required by an EFT can only be calculated from the lower level theory.  But once the LECs are determined and the desired accuracy specified, any description of the emergent phenomena with an
EFT stands on equal footing with the description using the lower level constituents.  We note that the EFT can operate by itself,
if one accepts to determine the LECs by a fit to data, that is with no recourse to the underlying theory.
 Any facts or predictions
(or collectively, \emph{truths}) obtained via the EFT, no matter how unexpected or seemingly disconnected
from the lower level point of view, is in principle deducible from the lower level domain.  Thus \emph{weak}
emergence~\cite{Chalmers2006-CHASAW} is automatically encompassed by EFTs. 

There is a plethora of examples where EFT predictions provided unexpected deeper insights into the workings
of lower level physics. A set of such EFT predictions refer to the chiral limit of QCD (i.e setting the light
quark masses to zero), where it can
be shown that certain quantities like the pion radius, the nucleon isovector radius or the electric and
magnetic polarizabilities of the nucleon diverge, see e.g.~\cite{Pagels:1974se,Bernard:1995dp}. Such a behavior
appears very unnatural (and can not be calculated in any way) from the point of view of QCD in terms of quarks
and gluons, but can be explained rather naturally in the EFT, where the Yukawa-suppressed pion cloud turns
into a quantity of infinite range  sampling all space and thus diverging. Another beautiful example is the
EFT investigation of the vailidity of carbon-oxygen based life on earth, where the amount of required fine-tuning
to keep the so-called Hoyle state in the spectrum of carbon-12 in close proximity to the three alpha-particle
threshold, thus enabling sufficient carbon and oxygen production in stars, under changes of the fundamental
parameters of the SM could be pinned down, see~\cite{Meissner:2014pma}  for details and more references.
Again, even though these predictions were made with EFTs operating at the higher levels, there is no question
that they can, \emph{in principle}, be deduced from lower-levels, but {\emph in practice}, this is currently
impossible and might be so for a long time.


%
%

\section{Anderson's ``More is different"\label{sect:more is different}}

Anderson's seminal paper ``More is different"~\cite{Anderson:1972pca} has been often used by philosophers and
scientists (mostly condensed matter physicists) alike as one of the main sources for the emergentist
``resurgence''~\cite{Mainwood2006}.  In his book \emph{How Can Physics Underlie the Mind}~\cite{ellis2016},
George Ellis appears to interpret Anderson's position as an anti-reductionist one (\cite{ellis2016}, pg. 4),
even misquoting Anderson's paper~\cite{Anderson:1972pca} on the very same page. However, Anderson, in his own
words, is a reductionist~\cite{Anderson:1972pca},
\\

\emph{As I said, we must all start with reductionism, which I fully accept.} (pg. 394),
\\

\noindent
which runs counter to the strong emergence mantra.  Anderson uses the term reductionism as synonymous with
what is referred to in metaphysics as microphysicalism~\cite{Mainwood2006}. However, the discrepancy between
Anderson's statement and Ellis' presentation of Anderson’s article~\cite{Anderson:1972pca} is more than a
liberal translation from one terminology into another.  The mismatch in terminology between Ellis and
Anderson becomes obvious in Anderson’s statement:
\\

\emph{(\ldots) the reductionist hypothesis does not by any means imply a constructionist one: The ability
  to reduce everything to simple fundamental laws does not imply the ability to start from those laws
  and reconstruct the universe.} (p. 393).
\\

\noindent
Anderson argues further that,  due to complexity, it may be impossible practically to compute the higher
level phenomena starting from the lower level constituents. He states,
\\

\emph{Surely there are more levels or organization between human ethology and DNA than there
  are between DNA and quantum electrodynamics, and each level can require a whole new conceptual
  structure.} (pg. 396).
\\

\noindent
This position alone does not imply anti-constructionism, because what can be meant by requiring a new
conceptual structure is the use of a new, more convenient basis for the description of relevant degrees
of freedom much along the lines of effective field theory. It is also not anti-reductionist, because
Anderson states that the novel concepts in physics are explained from fundamental laws (even when it takes
thirty years to do so, as in the case of superconductivity). However, the placement of this statement in
Anderson's paper (he used it to argue against an approach by some molecular biologists at the time ``to
reduce everything about the human organism to `only' chemistry, from the common cold and all mental disease
to the religious instinct'' (p. 396)) had likely attracted the interpreters in favour of strong emergence.

While it is clear that Anderson's position maintains that the new conceptual structure in complex systems
may not in practice be derived from the interaction of its constituents, interpreting it as an argument for
strong emergence would be equivalent to neglecting Anderson's full acceptance of reductionism. To ``start
with reductionism'' (pg. 394) matters here, even if one interprets Anderson's term ``reductionism''
as ``anti-constructionism''. Anderson at no point argues that the new conceptual structure of the higher
level of organization cannot be deduced from the lower-level constituents in principle.  On the other hand,
strong emergent phenomena are not deducible even in principle from the ``truths in the low-level
domain''~\cite{Chalmers2006-CHASAW}. Thus, Anderson's argumentation in~\cite{Anderson:1972pca} is
aligned with weak emergence.

%

\section{``Purpose" in life and physics\label{sect:purpose}}

In his book \emph{How Can Physics Underlie the Mind}~\cite{ellis2016} Ellis argues for strong emergence
in the case of consciousness and a number of other phenomena. For example, in the case of biology,
he argues that
\\

\emph{``Biology cannot be reduced to physics, because it has an ineliminable teleology component to its
  explanations. . ."} (pg. 373).
\\

\noindent
In other words, biology, and biological organisms by extension, have purpose, whereas physics does not.
Thus biology can not be reduced to physics.  He continues further,
\\

\emph{``Purposeful design underlies all the features we expect in life today (. . .).  But that
  physics knows nothing of these plans and theories."} (pgs. 414-415).
\\

\noindent
Again, the argument seems to be that since physics lacks the capacity to ``know'' the purpose of life,
strong emergence must exist. We find such statements reminiscent of the arguments made by proponents
of intelligent design (just replace the word \emph{Purposeful} with \emph{Intelligent} in the sentence
above)~\cite{NAP6024}.  We also caution in using terminology that may be precise at
one level, e.g. \emph{purpose}, but ill-defined at another level, as this ultimately adds confusion.

However, it may very well be that the ``purpose'' of some biological organism, seen from our limited
point of view, is procreation and the continuation of its species, but at the same time is equivalent
to the minimization of energy in some very complex phase space.  The latter explanation is fully consistent
with a physical interpretation of phenomena.  Neither our explanation nor Ellis' explanation is sufficient
to argue for one case or the other.  Our present ignorance of the  intricate workings of biological
organisms, or even the \emph{consciousness} of such beings, and its connection to physics is not a sufficient
argument for the existence of strong emergence.

Furthermore, we have not even come close to exhausting all possible research connections between physics
(and even lower levels) and biology. Bohr writes~\cite{heisenberg1996teil}
\\

\emph{``The richness of mathematical forms present in the quantum theory is perhaps by now sufficient
  to represent also biological forms."}
\\

\noindent
And, perhaps surprisingly here, Schr\"odinger makes a statement~\cite{Schrodinger1944-SCHWIL-5} which clarifies his position better
than Ellis' citation on pg.~4 of~\cite{ellis2016},
\\

\emph{We must therefore not be discouraged by the difficulty of interpreting life by the ordinary laws of
physics. For that is just what is to be expected from the knowledge we have gained of the structure of
living matter. We must be prepared to find a new type of physical law prevailing in it. Or are we to term
it a non-physical, not to say a super-physical, law? (\ldots) No. I do not think that. For the new principle
that is involved is a genuinely physical one: it is, in my opinion, nothing else than the principle of quantum
theory over again.}
\\

\noindent
The proposition that quantum mechanics plays an integral part in the emergence of life (and by extension,
conciousness) is evoked quite often, though admittingly we find such a connection hard to imagine.  But then
again, we are limited by our lack of imagination.

\section{Popper's falsifiability test\label{sect:popper}}

Any new scientific paradigm must be testable from a scientific point of view.  New (and even currently accepted)
theories make predictions which are subsequently tested empirically.  If the tests fail, then the theories
are either abandoned in favor of others or, more commonly, modified to better reflect reality.  Either
way, such testing provides better insight into Nature and her workings.  This process repeats itself over
and over, and is the basis for scientific progress.   

The great philosopher Sir Karl Popper argued that to be deemed ``scientific", a paradigm has to be
falsifiable~\cite{Keuth2019}.
The argument at first seems counter-intuitive, but Popper's own example in~\cite{Popper1962-POPCAR} serves as
an excellent explanation.  Popper, at the beginning of the 20th century, considered the scientific worth
of Einstein's theory of general relativity compared to Astrology.  Both made predictions:  Einstein's theory,
for example, predicted the bending of light from distant stars traversing near our
sun~\cite{eistein1915p1,eistein1915p2},
while astrologists predicted
marriage probabilities, lottery winnings, horse races, and so on, based off the alignment of planets
and other heavenly objects.  Einstein's theory could be easily falsified:  it made a prediction about
the exact amount of deflection about our sun of light originating from the Hyades star cluster
that could be directly empirically verified.  A false finding would directly lead to
abandonment.  On the other hand, false predictions by astrologists could always be argued away
(``Venus and Mars were aligned, but Jupiter was slightly askew $\ldots$").  History tells us
the rest~\cite{eddington1919}:
general relativity is one of mankind's greatest theories that has profoundly shaped our understanding of
space and time.  Sadly, astrology is still with us.

Using Popper's scrutiny the concept of strong emergence is \emph{not} ``scientific''.  We are not aware
of any \emph{predictions} its theories have made.  To be very clear on this issue, we refer to a prediction
by a quantifiable statement of a theory or model that is amenable to an experimental test, like the abovementioned
light bending in general relativity. Furthermore, if there are any predictions, it seems
that the mere complexity of the systems in which it is intended to be applied to leaves very little room
for direct falsifiability:  there is always some conditional statements which can be concocted (after
the fact) to ``argue away''  negative findings.  Under these circumstances strong emergence does not appear worthier
than astrology.

\section{Conclusion}
\label{sect:conc}
Effective field theories provide powerful tools for modern day physicists to understand and describe emergent
phenomena.  Though the lower-level theory can be preferable to work with since it can calculate certain quantities, e.g. LECs,
that are not accessible to an EFT, in practice calculations at the lower level are usually much more difficult or nearly
impossible.  Fortunately, once certain low-energy constants are determined from the lower-level theory or from data,
an EFT is an equally valid description of the higher-level
phenomena.  Furthermore, EFTs can, and \emph{do}, make predictions of the behavior of  emergent phenomena that can be
tested and falsified.  We have listed the necessary and sufficient conditions for the construction and
applicability of EFTs, and have argued that the principles underlying EFTs are fully consistent with the
\emph{methodological} and \emph{theoretical} reductionist point of view coupled with the weak form of emergence and causation.   

History is replete with examples of phenomena that have seemingly no connection to the laws of nature known
to man at that time.  A classic example is the formation of the rainbow~\footnote{For us, a rainbow is a physical
  phenomenon, it can be measured and artificially produced, and does not require the discussion of human impression. Or stated
  more simply: A rainbow is a rainbow is a rainbow.}.
  During biblical times, such a
phenomenon could surely be used as fodder for strong emergence.   But through the passage of time, as our
understanding of light refraction via water molecules has improved, our need to invoke strong emergence
disappeared and instead became weak emergence.   Indeed we can deduce the connection between light scattering
and the colorful arcs in the sky.  In this sense, strong emergent phenomena, if you will, are only
fleeting designators for unexplainable phenomena to be superseded by weak emergent phenomena as our
understanding of Nature improves.  

As the separation of scales between levels widens, it becomes invariably more difficult to see the
``connections'' between these levels.  Nothing annoys, yet motivates, physicists more (and presumably all
scientists and philosophers) than phenomena that are currently inexplicable with our current knowledge of
fundamental laws.  Indeed, such instances (e.g. dark matter and dark energy) hint at the possibility that
our current knowledge is insufficient, or our expected view of how nature works is too limited.  Most
likely it is a combination of both.  In these cases it is tempting to propose a new scientific paradigm
that somehow absolves us of our ignorances and lack of imagination, but to do so in a manner that is
not testable and verifiable inevitably does more harm (scientifically) than not.

\subsection*{Acknowledgements}

We thank Markus Gabriel for  helpful comments.  TL also thanks E. Berkowitz, C. Hanhart, J.-L. Wynen, and A. Wirzba for insightful discussions related
to this manuscript.  TL especially thanks M. Hru$\breve{\text{s}}$ka for her careful reading of
this manuscript and analysis. This work  was supported in part by the Deutsche
Forschungsgemeinschaft (DFG)  through funds provided to the Sino-German CRC~110
``Symmetries and the Emergence of Structure in QCD" (Grant No. TRR110),  by
the Chinese Academy of Sciences (CAS) through a President's International Fellowship
Initiative (PIFI) (Grant No. 2018DM0034) and by the VolkswagenStiftung (Grant No. 93562).



\begin{thebibliography}{99} 
  \frenchspacing


\bibitem{clay} {\tt http://www.claymath.org/millennium-problems}.  
  
\bibitem{Anderson:1972pca}
  P.~W.~Anderson,
  ``More Is Different,''
  Science {\bf 177} (1972)  393.
    
\bibitem{ellis2016}
George Ellis,
\newblock {\em How can physics underlie the mind? Top-down causation in the
 human context},
\newblock Springer-Verlag, 2016.
    
\bibitem{Heisenberg:1935qt}
  W.~Heisenberg and H.~Euler,
  ``Consequences of Dirac's theory of positrons,''
  Z.\ Phys.\  {\bf 98} (1936)   714.

\bibitem{Euler:1935zz}
  H.~Euler and B.~Kockel,
  ``Ueber die Streuung von Licht an Licht nach der Diracschen Theorie,''
  Naturwiss.\  {\bf 23} (1935) 246.

\bibitem{Karplus:1950zz}
  R.~Karplus and M.~Neuman,
  ``The scattering of light by light,''
  Phys.\ Rev.\  {\bf 83} (1951) 776.


\bibitem{Fermi:1934hr}
  E.~Fermi,
  ``An attempt of a theory of beta radiation. 1.,''
  Z.\ Phys.\  {\bf 88} (1934) 161-177
  

\bibitem{Weinberg:1978kz}
  S.~Weinberg,
  ``Phenomenological Lagrangians,''
  Physica A {\bf 96} (1979)  327.

\bibitem{Gasser:1983yg}
  J.~Gasser and H.~Leutwyler,
  ``Chiral Perturbation Theory to One Loop,''
  Annals Phys.\  {\bf 158} (1984) 142. 

\bibitem{Epelbaum:2008ga}
  E.~Epelbaum, H.~W.~Hammer and U.~G.~Mei{\ss}ner,
  ``Modern Theory of Nuclear Forces,''
  Rev.\ Mod.\ Phys.\  {\bf 81} (2009) 1773--1825

\bibitem{Lahde:2019npb}
  T.~A.~L\"ahde and U.-G.~Mei{\ss}ner,
  ``Nuclear Lattice Effective Field Theory : An introduction,''
  Lect.\ Notes Phys.\  {\bf 957} (2019) 1.

\bibitem{doi:10.1111/0029-4624.31.s11.17}
Mark~A. Bedau,
\newblock ``Weak emergence",
\newblock Nous, 31 {\bf s11} (1997) 375-399, 1997.

\bibitem{Mainwood2006}
Paul Mainwood,
\newblock {\em Is More Different? Emergent Properties in Physics},
\newblock PhD thesis, Merton College, University of Oxford, 2006.

\bibitem{Chalmers2006-CHASAW}
David~J. Chalmers,
\newblock ``Strong and weak emergence",
\newblock In P.~Davies and P.~Clayton, editors, {\em The Re-Emergence of
Emergence: The Emergentist Hypothesis From Science to Religion}, Oxford
University Press, 2006.

%
\bibitem{Pagels:1974se}
  H.~Pagels,
  ``Departures from Chiral Symmetry: A Review,''
  Phys.\ Rept.\  {\bf 16} (1975) 219.

\bibitem{Bernard:1995dp}
  V.~Bernard, N.~Kaiser and U.-G.~Mei{\ss}ner,
  ``Chiral dynamics in nucleons and nuclei,''
  Int.\ J.\ Mod.\ Phys.\ E {\bf 4} (1995) 193.

\bibitem{Meissner:2014pma}
  U.-G.~Mei{\ss}ner,
  ``Anthropic considerations in nuclear physics,''
  Sci.\ Bull.\  {\bf 60} (2015)  43. 

\bibitem{NAP6024}
National~Academy of~Sciences,
\newblock {\em Science and Creationism: A View from the National Academy of
Sciences, Second Edition},
\newblock The National Academies Press, Washington, DC, 1999.

\bibitem{heisenberg1996teil}
W.~Heisenberg,
\newblock {\em Der Teil und das Ganze: Gespr{\"a}che im Umkreis der
  Atomphysik},
\newblock Piper Taschenbuch, Piper, 1996.

\bibitem{Schrodinger1944-SCHWIL-5}
Erwin Schr\"odinger,
\newblock {\em What is Life? The Physical Aspect of the Living Cell},
\newblock Cambridge University Press, 1944.

\bibitem{Keuth2019}
Herbert Keuth,
\newblock {\em Karl Poppers „Logik der Forschung``}, pages 45--63,
\newblock Springer Fachmedien Wiesbaden, Wiesbaden, 2019.

\bibitem{Popper1962-POPCAR}
Karl Popper,
\newblock {\em Conjectures and Refutations: The Growth of Scientific
Knowledge},
\newblock Routledge, 1962.

\bibitem{eistein1915p1}
A.~Einstein,
``Grundgedanken der allgemeinen Relativit\"atstheorie und Anwendung dieser Theorie in der Astronomie",
Preussische Akademie der Wissenschaften, Sitzungsberichte, (1915) (part 1), 315

\bibitem{eistein1915p2}
A.~Einstein,
``Erkl\"arung der Perihelbewegung des Merkur aus der allgemeinen Relativit\"atstheorie",
Preussische Akademie der Wissenschaften, Sitzungsberichte, 1915 (part 2), 831–839

\bibitem{eddington1919}
A.~S.~Eddington,
``The total eclipse of 1919 May 29 and the influence of gravitation on light",
The Observatory {\bf 42} (1919) 119-122.

  
\end{thebibliography}
\end{document}